\documentstyle{article}

\title{\bf Formalizing the gene centered view of evolution}
\author{Y.\ Bar-Yam \\
New England Complex Systems Institute \\
24 Mt.\ Auburn St., Cambridge MA \\
Department of Molecular and Cellular Biology \\
Harvard University, Cambridge, MA}
\date{}

\begin{document}
\maketitle

\begin{abstract}
A historical dispute in the conceptual underpinnings of evolution is
the validity of the gene centered view of
evolution.~\cite{dawkins,lewontin} We transcend this debate by
formalizing the gene centered view as a dynamic version of the mean
field approximation. This establishes the conditions under which it is
applicable and when it is not. In particular, it breaks down for trait
divergence which corresponds to symmetry breaking in evolving
populations.
\end{abstract}

The gene centered view addresses a basic problem in the interplay of
selection and heredity in sexually reproducing organisms. Sexual
reproduction disrupts the simplest view of evolution because the
offspring of an organism are often as different from the parent as
organisms that it is competing against. In the gene centered view the
genes serve as indivisible units that are preserved from generation to
generation. In effect, different versions of the gene, i.e.\ alleles,
compete rather than organisms. It is helpful to explain this using the
``rowers analogy'' introduced by Dawkins.~\cite{dawkins} In this
analogy boats of mixed left- and right-handed rowers are filled from a
common rower pool. Boats compete in heats and it is assumed that a
speed advantage exists for boats with more same-handed rowers. The
successful rowers are then reassigned to the rower pool for the next
round. Over time, a predominantly and then totally single handed rower
pool will result. Thus, the selection of boats serves, in effect, to
select rowers who therefore may be considered to be competing against
each other. Other perspectives on evolution distinguish between
vehicles of selection (the organisms) and replicators (the genes). 
However, a direct analysis of the gene centered view to reveal its
domain of applicability has not yet been discussed. The analysis
provided here, including all the equations, is applicable quite
generally, but for simplicity it will be explained in terms of the
rowers analogy.\footnote{The rowers analogy may be considered a model
of a single gene in an $n$-ploid organism with $n$ the number of rowers,
or a model of $n$ genes with two alleles per gene and each pair labeled
correspondingly. The formal discussion applies to complete genomes i.e.\ 
to homolog and non-homolog genes.}

The formal question is: Under what conditions (assumptions) can
allelic (rower) competition serve as a surrogate for organism (boat)
competition in the simple view of evolution. Formalizing this question
requires identifying the conditions attributed to two steps in models
of evolution, selection and reproduction. In the selection step,
organisms (boats) are selected, while in the sexual reproduction step,
new organisms are formed from the organisms that were selected. This
is not fully discussed in the rowers model, but is implicit in the
statement that victorious rowers are returned to the rower pool to be
placed into new teams. The two steps of reproduction and selection can
be written quite generally as:
\begin{eqnarray}
\{N(s;t)\} & = & R[\{N'(s;t-1)\}] \\
\{N'(s;t)\} & = & D[\{N(s;t)\}]
\end{eqnarray}
\noindent The first equation describes reproduction. The number of
offspring $N(s;t)$ having a particular genome $s$ is written as a
function of the reproducing organisms $N'(s;t-1)$ from the previous
generation. The second equation describes selection. The reproducing
population $N'(s;t)$ is written as a function of the same generation
at birth $N(s;t)$. The brackets on the left indicate that each
equation represents a set of equations for each value of the genome. 
The brackets within the functions indicate, for example, that each of
the offspring populations depends on the entire parent population.

To formalize the gene centered view, we introduce a dynamic form of
what is known in physics as the mean field approximation. In the mean
field approximation the probability of appearance of a particular
state of the system (i.e.\ a particular genome, $s$) is the product of
probabilities of the components (i.e.\ genes, $s_i$)
\begin{equation}
P(s_1,\ldots,s_N) = \prod P(s_i).
\end{equation}
\noindent In the usual application of this approximation, it can be
shown to be equivalent to allowing each of the components to be placed
in an environment which is an average over the possible environments
formed by the other components of the system, hence the term ``mean
field approximation.'' The key to applying this in the context of
evolution is to consider carefully the effect of the reproduction
step, not just the selection step.

In many models of evolution that are discussed in the literature, the
offspring are constructed by random selection of surviving alleles (a
panmictic population). In the rowers analogy the return of successful
rowers to a common pool is the same approximation. This approximation
eliminates correlations in the genome that result from the selection
step and thus imposes Eq.\ (3), the mean field approximation,
on the reproduction step for the alleles of offspring. Even though it
is not imposed on the selection step, inserting this approximation
into the two step process allows us to write both of the equations in
Eq.\ (4) together as an effective one-step update
\begin{equation}
P'(s_i;t) = \tilde{D}[\{P'(s_i;t-1)\}]
\end{equation}
\noindent which describes the allele population change from one
generation to the next of offspring at birth. Since this equation
describes the behavior of a single allele it corresponds to the gene
centered view.

There is still a difficulty pointed out by Lewontin.~\cite{lewontin}
The effective fitness of each allele depends on the distribution of
alleles in the population. Thus, the fitness of an allele is coupled
to the evolution of other alleles. This is apparent in Eq.\ (4)
which, as indicated by the brackets, is a function of all the allele
populations. It corresponds, as in other mean field approximations, to
placing an allele in an average environment formed from the other
alleles. For example, there is a difference of likelihood of victory
(fitness) between a right-handed rower in a predominantly left-handed
population, compared to a right-handed rower in a predominantly
right-handed population. Since the population changes over time,
fitnesses are time dependent and therefore not uniquely defined. This
problem with fitness assignment would not be present if each allele
separately coded for an organism trait. While this is a partial
violation of the simplest conceptual view of evolution, however, the
applicability of a gene centered view can still be justified, as long
as the contextual assignment of fitness is included. When the fitness
of organism phenotype is dependent on the relative frequency of
phenotypes in a population of organisms it is known as frequency
dependent selection, which is a concept that is being applied to genes
in this context.

A more serious breakdown of the mean field approximation arises from
what is known in physics as symmetry breaking. This corresponds in
evolution to trait divergence of subpopulations. Such trait divergence
arises when correlations in reproduction exist so that reproduction
does not force complete mixing of alleles. The correlations in
reproduction do not have to be trait related. For example, they can be
due to spatial separation of organisms causing correlations in
reproduction among nearby organisms. Models of spatially distributed
organisms are sometimes called models of spatially structured
environments. However, this terminology suggests that the environment
itself is spatially varying and it is important to emphasize that
symmetry breaking / trait divergence can occur in environments that
are uniform (hence the terminology ``symmetry breaking''). In the rowers
model this has direct meaning in terms of the appearance of clusters
of mostly left and mostly right handed rowers if they are not
completely mixed when reintroduced and taken from the rower pool. 
Trait related correlations in sexual reproduction (assortive mating)
caused by, e.g.\ sexual selection, would also have similar consequences. 
In either case, the gene centered view would not apply.

Historically, the gene-centered view of evolution has been part of the
discussion of attitudes toward altruism and group selection and
related socio-political as well as biological concerns.~\cite{sober}
Our focus here is on the mathematical applicability of the
gene-centered view in different circumstances. While the formal
discussion we present here may contribute to the socio-political
issues, we have chosen to focus here on mathematical concerns.

The problem of understanding the mean-field approximation in
application to biology is, however, also relevant to the problem of
group selection. In typical models of group selection asexually
(clonally) reproducing organisms have fecundities determined both by
individual traits and group composition. The groups are assumed to be
well defined, but periodically mixed. Similar to the gene-centered
model, an assumption of random mixing is equivalent to a mean field
theory. Sober and Wilson (1998) have used the term ``the averaging
fallacy'' to refer to the direct assignment of fitnesses to
individuals. This captures the essential concept of the mean-field
approximation. However, both the limitations of this approximation in
some circumstances and its usefulness in others do not appear to be
generally recognized. For example, it is not necessary for well
defined groups to exist for a breakdown in the mean-field
approximation to occur. Correlations in organism influences are
sufficient. Moreover, standard group-selection models rely upon
averaging across groups with the same composition. For this case,
where well defined groups exist and correlations in mixing satisfy
averaging (mean-field) assumptions by group composition, equations
developed by Price\footnote{See discussion on pp.\ 73--74 of
\cite{sober}.} separate and identify both the mean field contribution
to fitness and corrections due to correlations. These equations do not
apply in more general circumstances when correlations exist in a
network of interactions and/or groups are not well defined, and/or
averaging across groups does not apply. It is also helpful to make a
distinction between the kind of objection raised by Lewontin to the
use of averaging, and the failure that occurs due to correlations when
the mean-field approximation does not apply. In the former case, the
assignment of fitnesses can be performed through the effect of the
environment influencing the gene, in the latter case, an attempt to
assign fitnesses to a gene would correspond to inventing non-causal
interactions between genes.

The mean field approximation is widely used in statistical physics as
a ``zeroth'' order approximation to understanding the properties of
systems. There are many cases where it provides important insight to
some aspects of a system (e.g.\ the Ising model of magnets) and others
where is essentially valid (conventional BCS superconductivity). The
application of the mean field approximation to a problem involves
assuming an element (or small part of the system) can be treated in
the average environment that it finds throughout the system. This is
equivalent to assuming that the probability distribution of the states
of the elements factor. Systematic strategies for improving the study
of systems beyond the mean field approximation both analytically and
through simulations allow the inclusions of correlations between
element behavior. An introduction to the mean-field approximation and
a variety of applications can be found in Bar-Yam
(1997).~\cite{baryam}

In conclusion, the gene centered view can be applied directly in
populations where sexual reproduction causes complete allelic mixing,
and only so long as effective fitnesses are understood to be relative
to the prevailing gene pool. However, structured populations (i.e.\
species with demes---local mating neighborhoods) are unlikely to
conform to the mean field approximation / gene centered view. 
Moreover, it does not apply to considering the consequences of trait
divergence, which can occur when such correlations in organism mating
occur. These issues are important in understanding problems that lie
at scales traditionaly between the problems of population biology and
those of evolutionary theory: e.g.\ the understanding of ecological
diversity and sympatric speciation.~\cite{sayama}

\end{document}